\documentclass[aip,amsmath,amssymb,reprint]{revtex4-1}

\usepackage{graphicx}
\usepackage{amsmath}
\usepackage[version=3]{mhchem} %
\usepackage{float}
\usepackage{adjustbox}

\newcommand{\kdotp}{$\mathbf{k}$$\cdot$$\mathbf{p}$}
\newcommand{\gapase}{GaP$_{1-x}$As$_{x}$}
\newcommand{\gapne}{GaP$_{1-x}$N$_{x}$}
\newcommand{\gaasne}{GaAs$_{1-x}$N$_{x}$}
\newcommand{\gapasne}{GaP$_{1-x-y}$As$_y$N$_x$}

\newcommand{\gapas}{Ga(PAs)}
\newcommand{\gapn}{Ga(PN)}
\newcommand{\gaasn}{Ga(AsN)}
\newcommand{\gapasn}{Ga(PAsN)}

\RequirePackage[normalem]{ulem}
\RequirePackage{color}\definecolor{RED}{rgb}{1,0,0}\definecolor{BLUE}{rgb}{0,0,1}

\begin{document}

\title[Electronic band structure of nitrogen diluted {\gapasn}]{Electronic band structure of nitrogen diluted {\gapasn}: formation of the intermediate band, direct and indirect optical transitions, localization of states}

\author{M. P. Polak}
\email[M. P. Polak: ]{maciej.polak@pwr.edu.pl}
\author{R. Kudrawiec}
\affiliation{Department of Experimental Physics, Faculty of Fundamental Problems of Technology, Wroclaw University of Science and Technology, Wyb. Wyspianskiego 27, 50-370 Wroclaw, Poland}
\author{O. Rubel}
\affiliation{Department of Materials Science and Engineering, McMaster University, 1280 Main Street West, Hamilton, Ontario L8S 4L8, Canada}

\date{\today}%

\begin{abstract}
The electronic band structure of Ga(PAsN) with a few percent of nitrogen is calculated in the whole composition of Ga(PAs) host using the state-of-the-art density functional methods including the modified Becke-Johnson functional to correctly reproduce the band gap, and band unfolding to reveal the character of the bands within the entire Brillouin zone. As expected, relatively small amounts of nitrogen introduced to Ga(PAs) lead to formation of an intermediate band below the conduction band which is consistent with the band anticrossing model, widely used to describe the electronic band structure of dilute nitrides. However, in this study calculations are performed in the whole Brillouin zone and reveal the significance of correct description of the band structure near the edges of Brillouin zone, especially for indirect band gap P-rich host alloy, which may not be properly captured with simpler models. The theoretical results are compared with experimental studies, confirming their reliability. The influence of nitrogen on the band structure is discussed in terms of application of Ga(PAsN) in optoelectronic devices such as intermediate band solar cells and light emitters. It is found that Ga(PAsN) with low N and As concentration has a band structure suitable for integration in Si tandem solar cells, since the lattice mismatch between Si and Ga(PAsN) is small in this case. Moreover, it is concluded that P-rich Ga(PAsN) alloys with low N concentration have a promising band structure for two colour emitters. Additionally, the effect of nitrogen incorporation on the carrier localization is studied and discussed.
\end{abstract}

\maketitle

\section{Introduction}\label{Ses:Introduction}
Nitrogen diluted III-V alloys are interesting to study in solid state physics for many reasons including the anticrossing interaction the in conduction band \cite{Shan19991221}, an enhanced conduction band non-parabolicity \cite{Lindsay1999443}, or carrier localization phenomena \cite{Kent20012613, Kudrawiec2013}, but one feature of these alloys makes them very unusual among other III-V alloys and thereby very interesting in applied physics. It is a simultaneous reduction of the band gap and the lattice constant due to the incorporation of nitrogen into III-V host. This feature makes dilute nitrides promising alloys for extending the emission of GaAs and InP-based lasers to longer wavelengths \cite{Bank2007773, Gladysiewicz2014996, Gladysiewicz2015} and desired light absorbers in multijunction solar cells \cite{Maros2016, Polojarvi2016, King2012801}. In addition dilute nitrides can be utilized as absorbers in intermediate band solar cells (IBSC), which have been proposed by \citet{Luque19975014} for obtaining high-efficiency single junction solar cells \cite{Luque2010160}. 

In recent years the concept of IBSC has been very intensively explored. Nowadays it is generally accepted that the intermediate band absorbers can be divided into three large groups: nanostructures, such as quantum dots \cite{Ramiro2014736, Okada2015, Sogabe2014, Liu2012237}, semiconductor bulk materials containing a high density of deep-level impurities \cite{Ramiro2014736, Okada2015, Luque2006320}, and highly mismatched alloys (HMAs) i.e., semiconductor alloys where the band anticrossing (BAC) effect takes place \cite{Shan19991221, Welna2015, Welna2014, Kudrawiec2014} and an intermediate band is formed.

Among HMAs, {\gapasn} with $\sim$40\% P and a few percent of nitrogen has been recognized as the most optimal for applications in IBSCs \cite{Kudrawiec2014}. Therefore in recent years this alloy has been intensively explored \cite{Kuang2013, baranowski_jap, Jussila2015, Ilahi2015291}. Moreover, {\gapasn} alloys with P concentration greater than 60\% have also been studied \cite{Biwa1998574, Robert2011, Babichev2014501, Zelazna2017}. This range of P concentration in {\gapasn} alloy seems to be the most interesting from the viewpoint of the intermediate band formation and the lattice matching with Si platform \cite{Zelazna2017}. The possible integration of {\gapasn} with Si platform is a very important advantage of this alloy, which also makes this material system very interesting for laser applications. For these purposes As-rich {\gapasn} alloys and quantum wells with a few percent of nitrogen atoms have been studied \cite{Kunert2006, Kunert2006361} and an electrically pumped laser on Si platform with the active region containing {\gapasn} quantum well has been demonstrated \cite{Liebich2011}. However further application of {\gapasn} alloy in IBSCs
and lasers integrated with Si platform needs better understanding of the electronic band structure for this alloy in the full Brillouin zone. It includes the understanding of formation of the intermediate band in this alloy, the role of localized and delocalized states, as well as the character of host band gap (direct vs indirect) with the change in the content of {\gapasn} alloy. 

To explain the composition dependence of the band gap in {\gapasn} the band anticrossing (BAC) model is very often utilized \cite{Kudrawiec2014, Kuang2013, baranowski_jap, Jussila2015}. According to this model the interaction of dispersionless N-related states with conduction band states of {\gapas} host is modelled as a two-level system by the following Hamiltonian \cite{Shan19991221,Wu2012SemicondSciTechn}
\begin{equation}\label{Eq:H-BAC}
H_{BAC}=
\left[
\begin{matrix}
& E_M(k) & C_{NM}\sqrt{x} \cr 
& C_{NM}\sqrt{x} & E_N
\end{matrix}\right]
\end{equation}
where $x$ is the mole fraction of substitutional N atoms and $C_{NM}$ is a coupling parameter which describes the interaction between the nitrogen level and the conduction band. This parameter depends on the semiconductor matrix and can be determined experimentally \cite{Shan19991221,Kudrawiec2014}. $E_M(k)$ is the energy dispersion of the lowest conduction band of the III-V matrix and $E_N$ is the energy of N-related states, all referenced to the top of the valence band. Eigenvalues $E_-(k)$ and $E_+(k)$ of the Hamiltonian (\ref{Eq:H-BAC}) form two highly non-parabolic subbands.
The $E_-$ band has features of a narrow intermediate band (IB) and is well separated by an energy gap from the $E_+$ band, i.e., the upper conduction band (CB), at the $\Gamma$-point. This gap is one of the subject of studies in this work since its existence in the \textit{full} Brillouin zone (BZ) is very important for the application of {\gapasn} as an IB absorber in IBSCs.

\begin{figure}[t]
\centering
\includegraphics{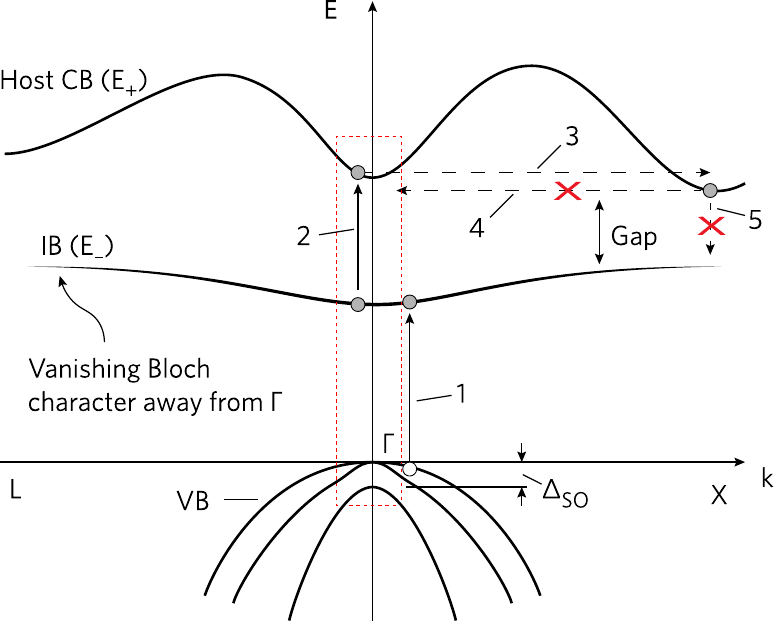}
\caption{Schematic band structure illustrating important ingredients and relevant processes in an intermediate band (IB) solar cell material of {\gapasn} family: 1 -- primary excitation, 2 -- secondary excitation, 3 -- scattering from $\Gamma$ to X valley, 4 -- backscattering, and 5 -- re-trapping. The direct excitation from the valence band (VB) to the host conduction band (CB) is not shown for simplicity. The doted red line shows a region of applicability for a combined {\kdotp} and BAC model.}
\label{Fig:IBSC-schematic}
\end{figure}

A HMA absorber in IBSC should utilize three bands resulting in the optical transitions from VB to CB ($\Gamma_v-E_+$ band), from VB to IB ($\Gamma_v-E_-$ band), and IB to CB, as schematically shown in Fig.~\ref{Fig:IBSC-schematic}. An optimal total band gap of the IBSC should be 1.95~eV (100\%) split in the ratio of approximately 36\%/64\% between the two optical transitions $E_- -E_+$ and $\Gamma_v-E_-$~\cite{Luque_NP_6_2012}. It is favourable for the CB to show a slight indirect character to scatter  electrons from $\Gamma$ valley and prolong their lifetime. In order to suppress carrier relaxation from CB to IB (i.e., to fulfil the condition necessary for an efficient IBSC operation) these bands have to be separated by an energy gap in the \textit{whole} BZ \cite{Luque19975014}. Such a separation exists within the BAC model, but is confirmed experimentally only at the \textit{center} of the Brillouin zone by modulation spectroscopy \cite{Shan19991221,Kudrawiec2014} since this technique probes only direct optical transitions  \cite{Shan19991221,Kudrawiec2014}. Therefore the question of existence of an energy gap between IB and CB in {\gapasn} and other HMAs is still open.

A combination of the {\kdotp} and BAC model is a popular choice for modelling the band structure of HMAs \cite{Buckers2010, Gladysiewicz2014996, Gladysiewicz2015}. However, the model applies only to electronic states in vicinity of $\Gamma$ point and ignores the disorder in IB, e.g. the degree of localization of electronic states, which is responsible for inhomogeneous broadening of a luminescence spectrum and a gain spectrum in lasers. Extending the conclusions drawn from the BAC and {\kdotp} approach to the electronic band structure in the full BZ is very interesting \cite{Kudrawiec2014,Heyman2017}, but can be controversial because of limitations of the {\kdotp} method. Therefore a proper modelling of the electronic band structure of {\gapasn} in the full BZ is very important taking into account the application of this alloy in IBSC as well as other devices. So far the electronic band structure of ternary Ga(NAs) and Ga(NP) alloys (i.e., extreme host content cases of {\gapasn}) has been calculated using empirical pseudopotential \cite{Bellaiche199710233,Kent20011152081}, tight-binding methods \cite{OReilly2004}, and the density functional theory (DFT) approaches \cite{Neugebauer1995,Sakamoto2014} but the issue of IB formation and the energy separation between the IB and the CB as well at the degree of localization of electronic states was not addressed in these alloys. The quaternary dilute nitride {\gapasn} have also not been previously studied by DFT methods in the full composition range.

The aim of this paper is to perform  calculations of the electronic band structure using  DFT based methods for Ga(NPAs) alloy in the full BZ for a full composition range of {\gapas} host and discuss the electronic band structure in the context of the application of Ga(NPAs) in IBSC and light emitters including the concept of two color emitters recently proposed for HMAs \cite{Welna2017}. We have applied state-of-the-art DFT methods together with a band unfolding technique based on spectral weights \cite{Popescu2010, Popescu2012}. It allows to obtain an effective band structure of a solid solution from folded supercell bands giving the possibility of distinguishing between direct and indirect gaps as well as provides insight into a Bloch character of the bands. To investigate the degree of localization of electronic states, an inverse participation ratio method \cite{Pashartis2017} has been applied. Since the main subject of this study is the quaternary {\gapasn} alloy, for additional credibility and broader extent and analysis of the results, all connected ternary alloys have been studied using the same method as well, i.e. the {\gapas} host in the whole composition range, and the {\gapn} and {\gaasn} alloys with N content limited up to around 10\% due to challenges associated with their practical synthesis.

\section{Computational methods}

The first-principles calculations have been carried out using density functional theory \cite{Hohenberg1964,Kohn1965} and the linear augmented plane wave method implemented in the \texttt{WIEN2k} package \cite{Blaha2001}. The SCAN functional \cite{Sun2016} has been used to capture exchange-correlation effects in the geometry optimization. The modified Becke-Johnson exchange potential  (mBJ) \cite{Tran2009} combined with the local density approximation (LDA) for correlation \cite{Perdew1992} has been used in the band structure calculations to obtain realistic band gaps \cite{Koller2011, Kim2010, Camargo2012}. Even though mBJLDA significantly improves over standard DFT semi-local exchange-correlation functionals for the description of band gaps, we  slightly modified the $c$ parameter to achieve a perfect agreement of the calculated direct band gaps with experimental (0~K) values for the parent binary compounds, since they are a starting point of all subsequent calculations. Table~\ref{Table:band-gaps} presents a comparison of our results for the band gaps with experimental values and lists the adjusted values of the $c$ parameter.

\begin{table*}[th!]
\caption{Calculated and experimental \cite{Vurg} values of band gaps and lattice constants $a$ for parent zinc-blend binary compounds; $c$ is the adjusted mBJLDA parameter. All values correspond to 0~K temperature except for GaN where the experimental value of $a=4.5$~{\AA} corresponds to 300~K.}\label{Table:band-gaps}
\begin{ruledtabular}
\begin{tabular}{lcccccc}
Property & \multicolumn{2}{c}{GaP} & \multicolumn{2}{c}{GaAs} & \multicolumn{2}{c}{GaN} \\
  & \multicolumn{2}{c}{($c=1.50$)} & \multicolumn{2}{c}{($c=1.53$)} & \multicolumn{2}{c}{($c=2.10$)} \\
  \cline{2-3}\cline{4-5}\cline{6-7}
 & DFT & Exp. & DFT & Exp. & DFT & Exp. \\ \hline
Direct gap (eV) & 2.89 & 2.89 & 1.52 & 1.52 & 3.30 & 3.30 \\
Indirect gap (eV) & 2.33 (X) & 2.35 (X) & 1.72 (L) & 1.81 (L) & 4.99 (X) & 4.52 (X) \\ 
$a$ (\AA) & 5.48 & 5.45 & 5.68 & 5.65 & 4.46 & 4.5
\end{tabular}
\end{ruledtabular}
\end{table*}

Solid solutions have been simulated via constructing 128-atom supercells by translating the 2-atom basis along the primitive lattice vectors of the zinc-blend structure with the multiplicity of $4\times4\times4$. A consistent way of distributing different elements within the supercell has been achieved by generating special quasirandom structures (SQS) \cite{Zunger1990} with the use of the \texttt{mcsqs} code \cite{VanDeWalle2013}. Quadruplets and triplets with the closest neighbor of the same kind as well as pairs to three closest neighbors have been considered for correlation functions, and the Monte Carlo algorithm has been run for $3\cdot 10^6$ iterations which has been found to be sufficient to either find a perfect match or for an objective function to converge.

Radii $R$ of muffin-tin spheres of As, Ga, P and N has been chosen to be 2.2, 2.2, 1.9 and 1.7~a.u. respectively. Considering the small muffin-tin sphere of the nitrogen atom and the size of a 128-atom supercell, convergence studies lead to a $R_\text{min}K_\text{max}=7$ for nitrogen-containing structures. The increased value of $R_\text{min}K_\text{max}=7.5$ was used for nitrogen-free GaP$_{1-x}$As$_{x}$ structures. The BZ of supercells was sampled with a Monkhorst-Pack mesh \cite{Monkhorst_PRB_13_1976} of $2\times2\times2$ (equivalent to $8\times8\times8$ in the primitive zinc blende unit cell).

Lattice parameters and the $c$-mBJLDA parameters of alloys were interpolated from binary solids. The geometry optimization procedure has been performed until the maximum forces acting on an atom did not exceed 1~mRy/Bohr with a convergence of 0.1~mRy/Bohr. In the band structure and density of states (DOS) calculations energies have been converged down to 1~mRy. For the DOS calculations in 128-atom supercells (Fig. \ref{Fig:gapnas_dos}), which have been crucial in determining the intermediate gap, a much denser $k$ mesh of $8\times8\times8$ (equivalent to $32\times32\times32$ in the primitive zinc blende unit cell) has been used together with an improved tetrahedron method \cite{Blochl1994} to achieve a better energy resolution and avoid false conclusions. Considering difficulties in interpretation of a band structure of supercells, especially in materials with an indirect band gap (GaP) as well as the formation of the intermediate band, the band unfolding technique described in \cite{Popescu2010, Popescu2012} and implemented as the \texttt{fold2Bloch} code \cite{Rubel2014} has been used.

\section{Results and discussion}
\subsection{Formation of the intermediate band}

As emphasized in Sec.~\ref{Ses:Introduction}, the properly calculated and unfolded DFT band structures provide us with a complete, whole BZ, picture of the energy band dispersions. This is particularly useful when the host material, {\gapas}, has an indirect gap up to around 51\% of its As composition range on the P rich side.

Figure~\ref{ALL_ALLOYS} (a)-(f) shows the unfolded band structures of six compositions of the GaP$_{1-x}$As$_{x}$ alloy, which will serve as a reference for {\gapasn}. The unfolded band structures throughout the whole composition range consist of eigenvalues with high Bloch spectral weights, mostly over 80\% at the band edges. This material is well known experimentally \cite{Vurg}, and its excellent optical properties have lead to its frequent use in light emitting diodes or multi-junction III-V solar cells. Those high spectral weights directly relate to the good optical quality of the material and a very low amount of localized states, which is discussed later in Sec.~\ref{Subsec:Carrier localization}. 
\begin{widetext}
\begin{turnpage}
\begin{figure}[H]
\centering
\includegraphics[width=1.37\textwidth]{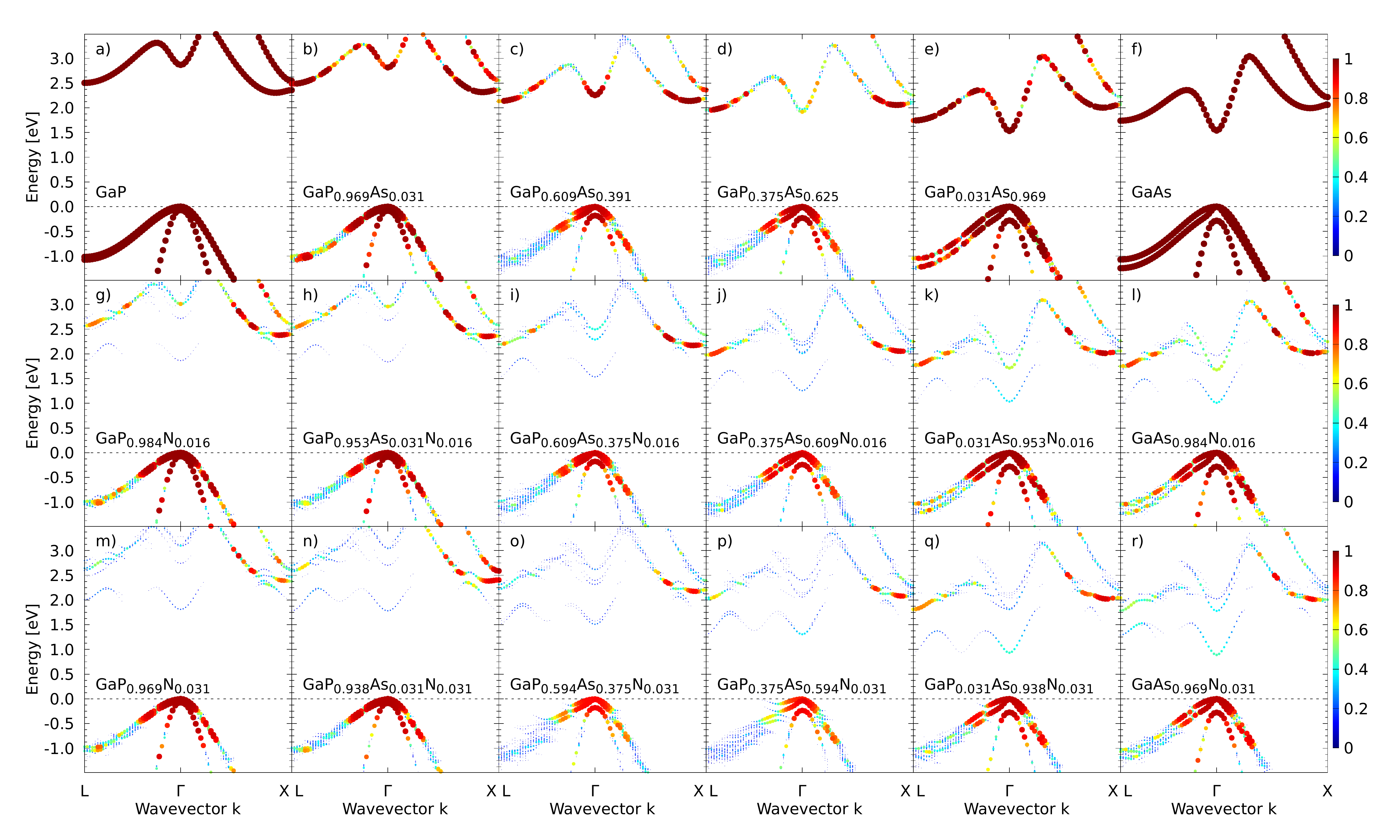}
\caption{Unfolded band structures of all the studied alloys organized row- and column-wise. P/As ratio changes from left to right (As increases), nitrogen concentration increases from top to bottom.}
\label{ALL_ALLOYS}
\end{figure}
\end{turnpage}
\end{widetext}

\begin{figure}[H]
\centering
\includegraphics[width=0.45\textwidth]{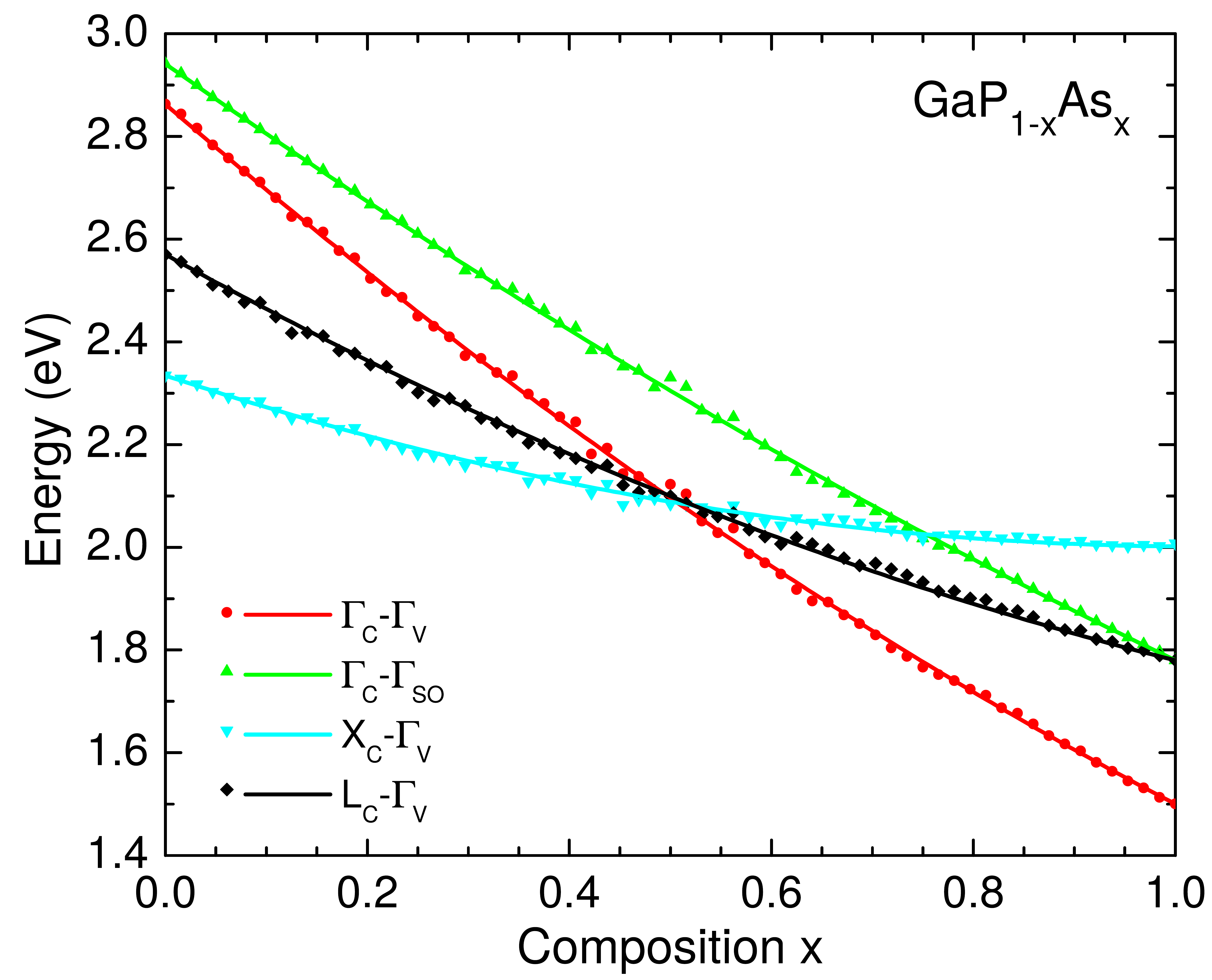}
\caption{Direct ($\Gamma_\text{v}\!-\!\Gamma_\text{c}$) and indirect ($\Gamma_\text{v}\!-\!L_\text{c}$ and $\Gamma_\text{v}\!-\!X_\text{c}$) band gaps in the host {\gapas} alloy as a function of As concentration. The points represent results from DFT calculations while the curves are fitted with Eq.~(\ref{vegard}) using parameters listed in Table~\ref{Table:Parameters}.}
\label{GaPAs_x}
\end{figure}

Figure~\ref{GaPAs_x} shows the energies of direct and indirect gaps as well as the spin-orbit split off band gap in {\gapas}. The points represent the values of energy gaps obtained from band edges of unfolded band structures and the lines are second order polynomial fits according to the Vegard's law with a constant bowing parameter $b$:
\begin{equation}
E_g^{AB_{1-x}C_x}=(1-x)E_g^{AB}+xE_g^{AC}-bx(1-x)
\label{vegard}
\end{equation}
where $E_g$ is the band gap and $x$ is the fractional composition. This is the widely used approximation of the composition dependence of the band gap used for regular semiconductor alloys.

The distinction between direct and indirect gaps has been enabled due to the band unfolding and is otherwise obstructed by zone folding in supercell calculations. It is important to notice that the lines in Fig.~\ref{GaPAs_x} provide a nearly perfect fit to the obtained data points. The scattering of $E_g(x)$ values is very small even near to the middle of the composition range. This behaviour is a consequence of a well-preserved Bloch character at the band edges. It is also a characteristic of an alloy without substantial localization of states resulting in desirable transport properties that is crucial for solar cell absorber materials.

The intersection of $\Gamma_\text{v}\!-\!\Gamma_\text{c}$ and $\Gamma_\text{v}\!-\!X_\text{c}$ gaps, i.e., the indirect-direct transition in Fig.~\ref{GaPAs_x}, occurs at around 51\% of As content. The bowing parameter for the direct band gap is $b=0.33$~eV (Table~\ref{Table:Parameters}), which is close to the bowing parameter $b=0.19$~eV recommended by Vurgaftman et al. \cite{Vurg}. This difference can be attributed to a finite temperature at which the experimental parameter has been obtained in contrast to our 0~K calculations. This discrepancy results in a maximum error of $35$~meV for the band gap at $x=0.5$ of this alloy and smaller differences at other alloy contents.  Therefore it can be assumed that our theoretical predictions are consistent with previous studies for this alloy \cite{Vurg} and ensure a solid starting point for calculations of the alloys with nitrogen.

\begin{table*}[t]
\caption{Parameters used in description of composition-dependent band gap of the studied alloys.}\label{Table:Parameters}
\begin{ruledtabular}
\begin{tabular}{cclcccc}
Equation & Transition & Parameter & {\gapase} & {\gapne} & {\gaasne} & {\gapasne} \\
\hline
 (\ref{vegard}) & $\Gamma_V-\Gamma_C$ & $b$ (eV) & 0.33 & $-$ & $-$ & 0.33 \\
                &  & $E_{AB}$ (eV) & 2.87 & $-$ & $-$ & $\frac{1}{2}\left(2.87+2.18\pm\sqrt{\left(2.87-2.18\right)^2+4\cdot3.05^2x}\right)$ \\
                &  & $E_{AC}$ (eV) & 1.52 & $-$ & $-$ & $\frac{1}{2}\left(1.52+1.65\pm\sqrt{\left(1.52-1.65\right)^2+4\cdot2.7^2x}\right)$ \\
                & $\Gamma_V-L_C$ & $b$ (eV) & 0.31 & $-$ & $-$ & 0.31 \\
                &  & $E_{AB}$ (eV) & 2.5 & $-$ & $-$ & $\frac{1}{2}\left(2.5+2.18+\sqrt{\left(2.5-2.18\right)^2+4\cdot1.65^2x}\right)$ \\
                &  & $E_{AC}$ (eV) & 1.72 & $-$ & $-$ & $\frac{1}{2}\left(1.72+1.65+\sqrt{\left(1.72-1.65\right)^2+4\cdot0.85^2x}\right)$ \\
                & $\Gamma_V-X_C$ & $b$ (eV) & 0.32 & $-$ & $-$ & 0.32 \\ 
                &  & $E_{AB}$ (eV) & 2.33 & $-$ & $-$ & $\frac{1}{2}\left(2.33+2.18+\sqrt{\left(2.33-2.18\right)^2+4\cdot0.97^2x}\right)$ \\
                &  & $E_{AC}$ (eV) & 1.99 & $-$ & $-$ & $\frac{1}{2}\left(1.99+1.65+\sqrt{\left(1.99-1.65\right)^2+4\cdot0.68^2x}\right)$ \\

\hline
 (\ref{bac_vegard}) & $E_\pm(\Gamma)$ & $E_M(\Gamma)$ (eV) & $-$ & 2.87 & 1.52  & $-$ \\ 
                    &              & $C_{NM}(\Gamma)$ (eV) & $-$ & 3.05 & 2.7  & $-$ \\
                    & $E_+(L)$   & $E_M(L)$ (eV)         & $-$ & 2.5 & 1.72 & $-$ \\
                    &              & $C_{NM}(L)$ (eV)      & $-$ & 1.65 & 0.85  & $-$ \\
                    & $E_+(X)$   & $E_M(X)$ (eV)         & $-$ & 2.33 & 1.99 & $-$ \\
                    &              & $C_{NM}(X)$ (eV)      & $-$ & 0.97 & 0.68 & $-$ \\ \cline{2-3}
                    &              & $E_N$ (eV)            & $-$ & 2.18 & 1.65 & $-$ \\
\end{tabular}
\end{ruledtabular}
\end{table*}

Before proceeding to the quaternary {\gapasn} alloy, it is worth establishing that properties of the ternary alloys {\gapn} and {\gaasn} can also be reproduced. Figure~\ref{ALL_ALLOYS} (g) and (m) shows the unfolded band structures for the {\gapn} alloy in the dilute-N regime. This material is one of the key representatives of HMAs, and its behavior is expected to be much different from an ordinary semiconductor alloy such as the previously discussed {\gapas}. For the low nitrogen content, a clearly visible intermediate band is formed. A decomposition of the DOS on different atoms and their orbitals reveals that the part of the DOS corresponding energetically to the intermediate band consists primarily of N-$s$ states, which has been expected from the BAC model.

\begin{figure}[H]
\centering
\includegraphics[width=0.4\textwidth]{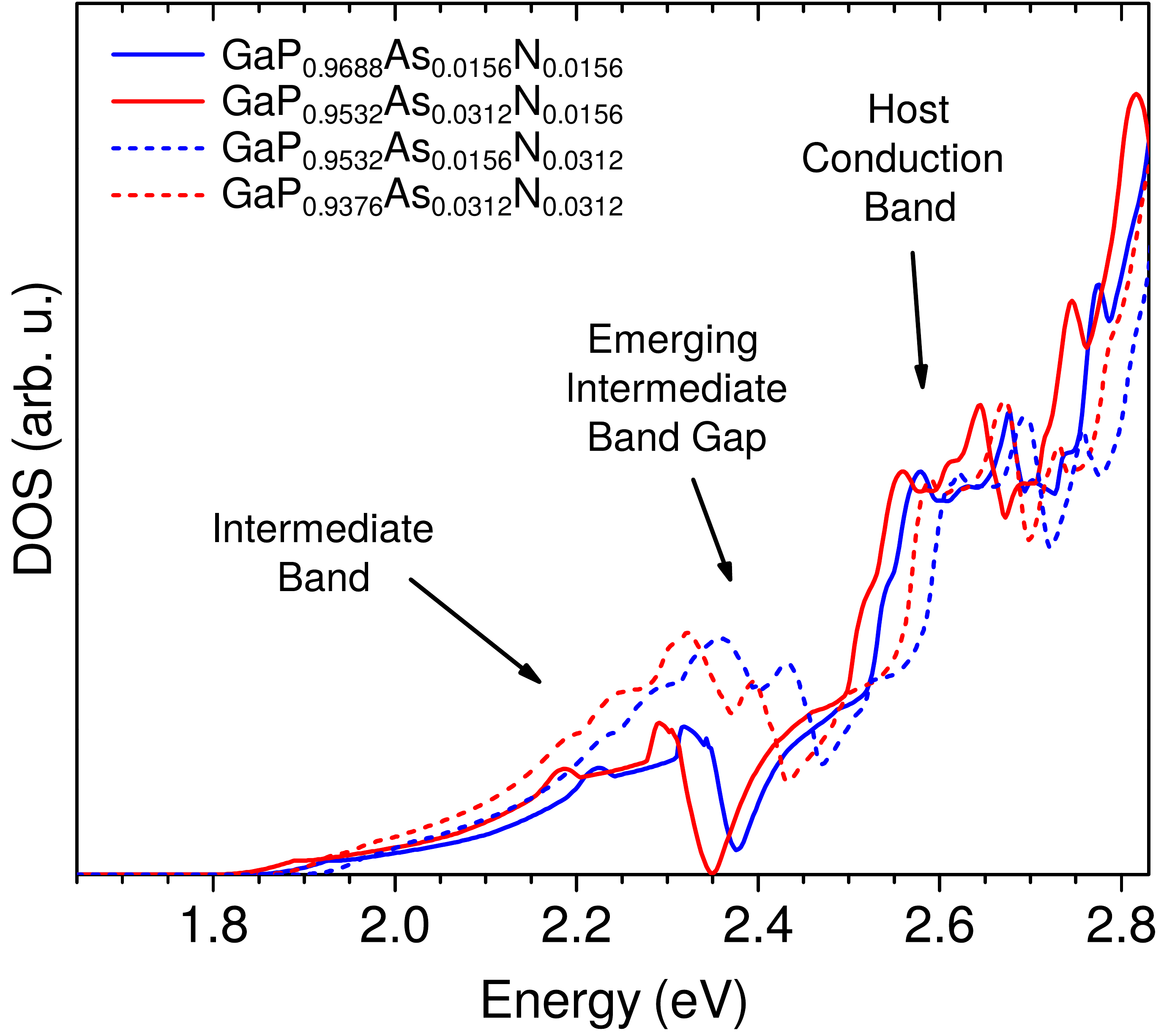}
\caption{Density of states for different As and N concentration. Solid and dashed lines correspond to 1.56\% and 3.12\% of nitrogen respectively, blue and red lines correspond to 1.56\% and 3.12\% of arsenic respectively.}
\label{Fig:gapnas_dos}
\end{figure}

It is difficult to confidently establish from the band structure alone a composition limit in which the energy gap between the nitrogen-induced IB and the host conduction band exists. The existence of this gap is an important ingredient for IBSC absorber materials (Fig.~\ref{Fig:IBSC-schematic}). For the lowest available composition in our 128-atom supercell, i.e. 1.56\% of N, the band structure suggest that a presence of such gap is possible (Fig.~\ref{ALL_ALLOYS} (g)), but as the nitrogen content increases, the gap seems to be closing. The gap can be properly identified on a well-resolved DOS constructed using a dense $k$ mesh, which is done for the {\gapasn} alloy (Fig.~\ref{Fig:gapnas_dos}) and will be discussed later.

\begin{figure}[H]
\centering
\includegraphics[width=0.45\textwidth]{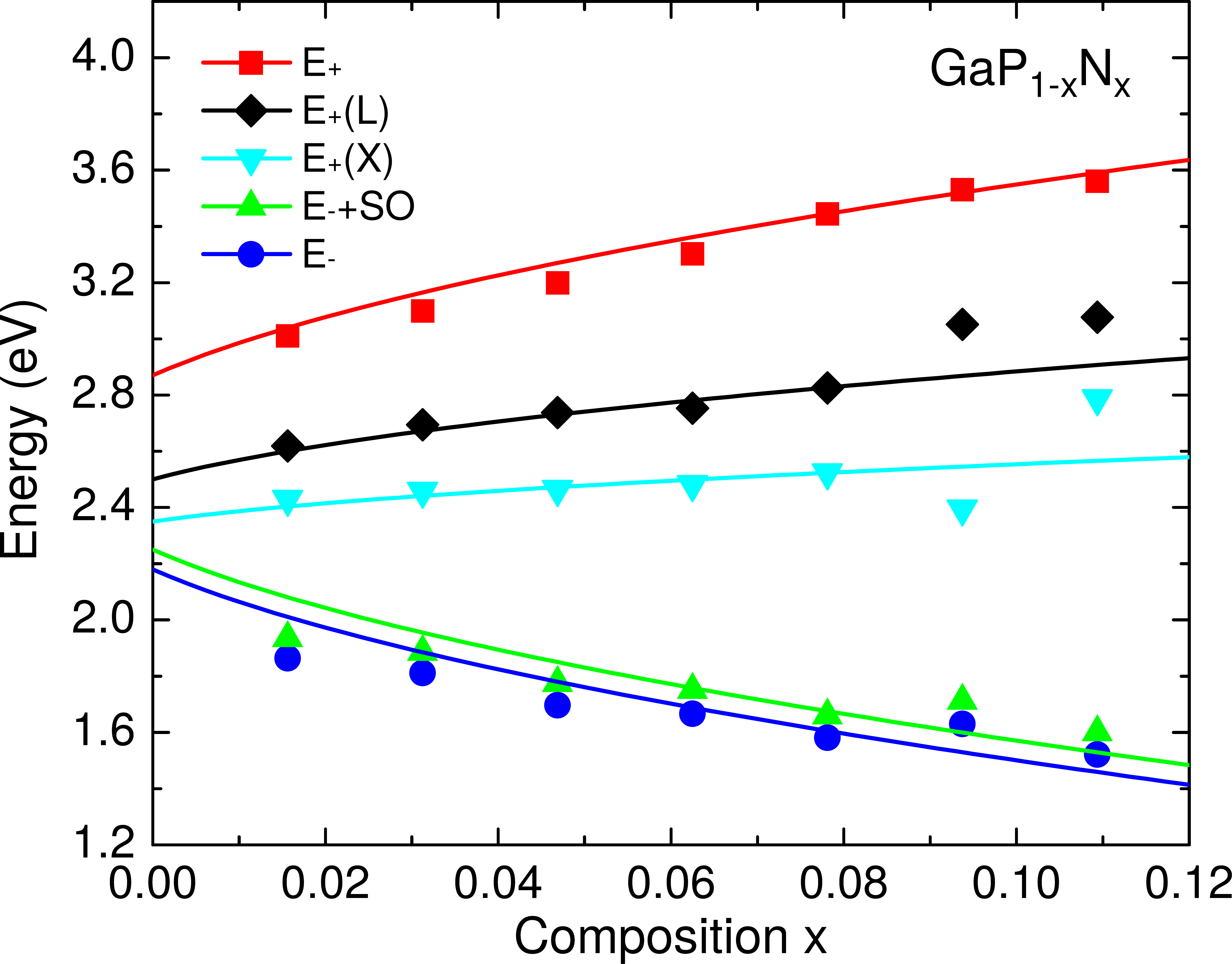}
\caption{Direct ($\Gamma$-$E_-$ and ($\Gamma$-$E_+$)) and indirect ($\Gamma$-$E_+(X)$ and $\Gamma$-$E_+(L)$) band gaps in the {\gapne} alloy as a function of N concentration. The points represent results of DFT calculations while the curves are fitted with Eq.~(\ref{bac_vegard}) using parameters listed in Table~\ref{Table:Parameters}.}
\label{GaPN_x}
\end{figure}

The energies of direct optical transitions inferred from unfolded band structures of GaP$_{1-x}$N$_x$ with $x<0.11$ are shown in Fig.~\ref{GaPN_x}. Since it is a HMA, the standard equation for the composition dependence of the band gaps [Eq.~(\ref{vegard})] could not be used to fit the curves. Instead, the BAC model is used [Eq.~(\ref{bac_vegard})] with parameters listed in Table~\ref{Table:Parameters}
\begin{equation}
\begin{split}
&E_{\pm}(k)=\\&=\frac{1}{2}\left\{E_M(k)+E_N\pm\sqrt{\left[E_M(k)-E_N\right]^2+4\left[C_{NM}(k)\right]^2x}\right\}
\end{split}
\label{bac_vegard}
\end{equation}

As a result of calculating the band structure within the whole BZ, a modification of the host conduction band due to the introduction of nitrogen can be studied not only in the $\Gamma$ point but also in the $L$- and $X$-points. The energy levels at the $L$ and $X$ points are less affected than at $\Gamma$, which reflects in lower values of the corresponding coupling parameters $C_{NM}(k)$ (Table~\ref{Table:Parameters}). Higher N-concentrations distort the host CB much more, yet still affect the center of the BZ the most. According to our DFT calculations the $s$-like character of the conduction band in GaP host is weaker at the end of BZ and therefore the $C_{NM}(k)$ element is smaller at the end of BZ. This agrees with the experimental results from \cite{Wu20022413031}, where the BAC interaction is found to be $k$ vector dependent, and its strength decreases away from the center of the BZ.

A very similar discussion of the band structure can be held for {\gaasn}, where a nitrogen band in a low concentration regime is visible on the unfolded band structures as well [Fig.~\ref{ALL_ALLOYS} (l) and (r)]. Here, however, the curvature of the intermediate band is more pronounced and the IB overlaps with the conduction band, eliminating the possibility of the material to have a well defined intermediate gap between the two bands (Fig. \ref{Fig:IBSC-schematic}).

\begin{figure}[H]
\centering
\includegraphics[width=0.45\textwidth]{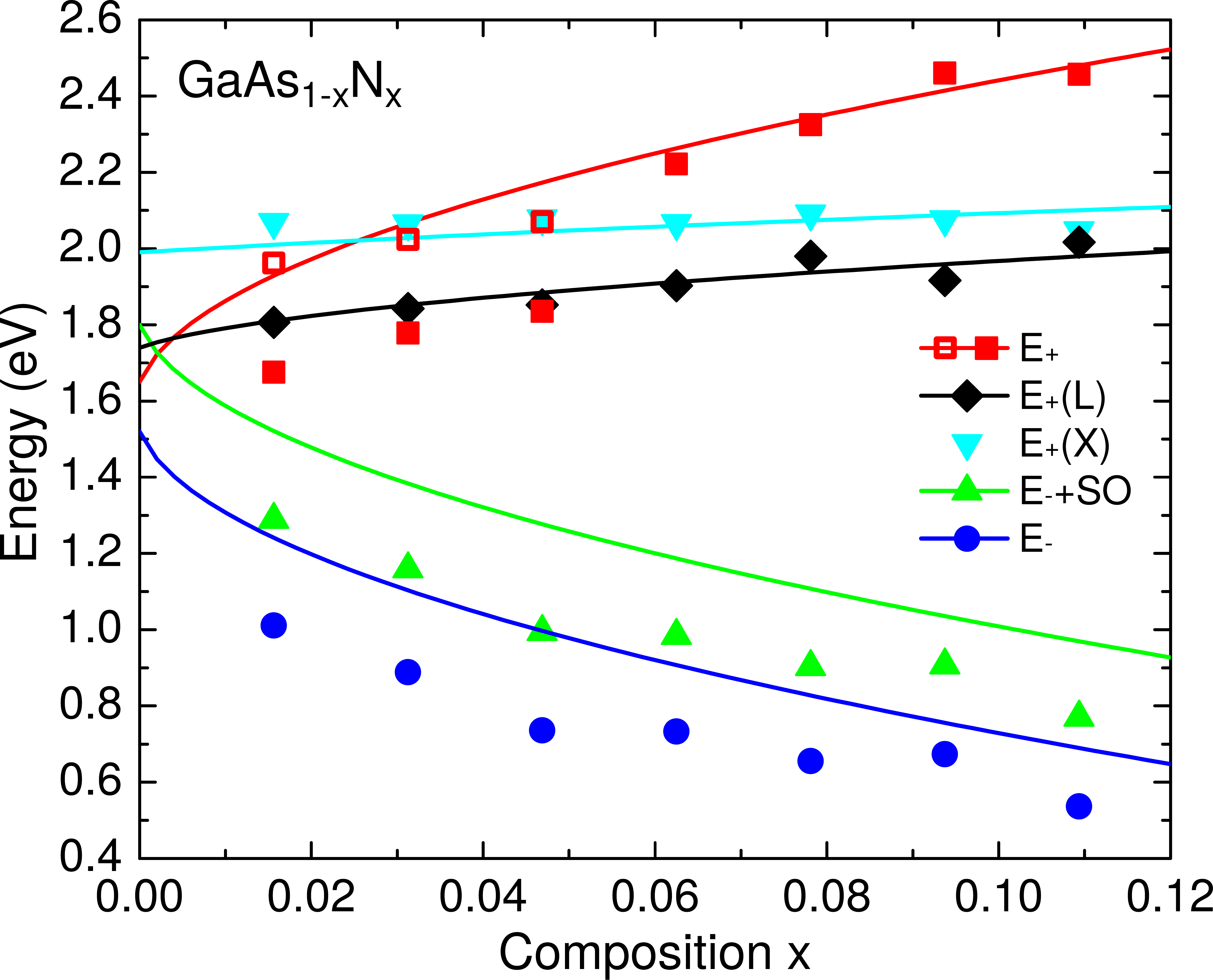}
\caption{Direct ($\Gamma$-$E_-$ and $\Gamma$-$E_+$) and indirect ($\Gamma$-$E_+(X)$ and $\Gamma$-$E_+(L)$) band gaps in the {\gaasne} alloy as a function of N concentration. The points represent results from DFT calculations while the curves are fitted with Eq.~(\ref{bac_vegard}) using parameters listed in Table~\ref{Table:Parameters}.}
\label{GaAsN_x}
\end{figure}

The energies of direct and indirect optical transitions obtained from unfolded band structures of GaAs$_{1-x}$N$_x$ with $x<0.11$ are shown in Fig.~\ref{GaAsN_x}. In this case, if the CB is split into two bands [Fig. \ref{ALL_ALLOYS} (l)] the lower spectral weight eigenvalue seems to follow the BAC curve much better. This splitting is shown in Fig.~\ref{GaAsN_x} as filled and open red squares. This artefact may result from a limited size of the supercell and periodical boundary conditions or proximity of N atoms. The wave-vector dependence of the coupling parameters $C_{NM}(k)$ has the same character as in GaP host, due to the similar $k$-vector dependence of the $s$-like character of the conduction band, although the magnitude of the dependence is different. In general, different $k$-vector dependence of the coupling parameter should be expected for different III-V hosts due do the different orbital structure and wavefunction character of the bands when moving from the center to the end of the BZ.

Results for low N concentrations in {\gapn} and {\gaasn} in combination with the composition dependence of the band gap of {\gapas} lead to a conclusion that a properly engineered {\gapasn} alloy should allow for a range of compositions of As and N where the intermediate band behaviour of {\gapn} and {\gaasn} can be combined with the possibility of the band gap tuning of {\gapas} as well as a possible integration with the Si platform by matching the lattice parameters. Our results for ternary alloys suggest that the intermediate band gap, desirable for solar cell applications, most likely exists only for a very low nitrogen content on the P-rich side of the {\gapasn} alloy.

Following these assumptions, an electronic band structure of the {\gapasn} alloy was studied for two lowest nitrogen concentrations, namely 1.56\% and 3.12\%, and a whole P/As composition range. Taking into account the fact that results of our effective band structure calculations match the experimental trends as well as the BAC model for the well known {\gapas} and low nitrogen concentration {\gapn} and {\gaasn} alloys, we expect that the methods used here should give reliable results for the band structure of the {\gapasn} alloy.

The effective band structures for selected compositions are presented in Fig.~\ref{ALL_ALLOYS}(n)--(p). As expected, the addition of nitrogen resulted in splitting of the conduction band into two $E_-$ and $E_+$ subbands. The disorder in the band structure increases for high P compositions of the host. States at the bottom of the IB become progressively  more $\Gamma$-like as the composition shifts toward the As-rich limit. The IB is always clearly visible and distinguishable. 

\begin{figure}[H]
\centering
\includegraphics[width=0.45\textwidth]{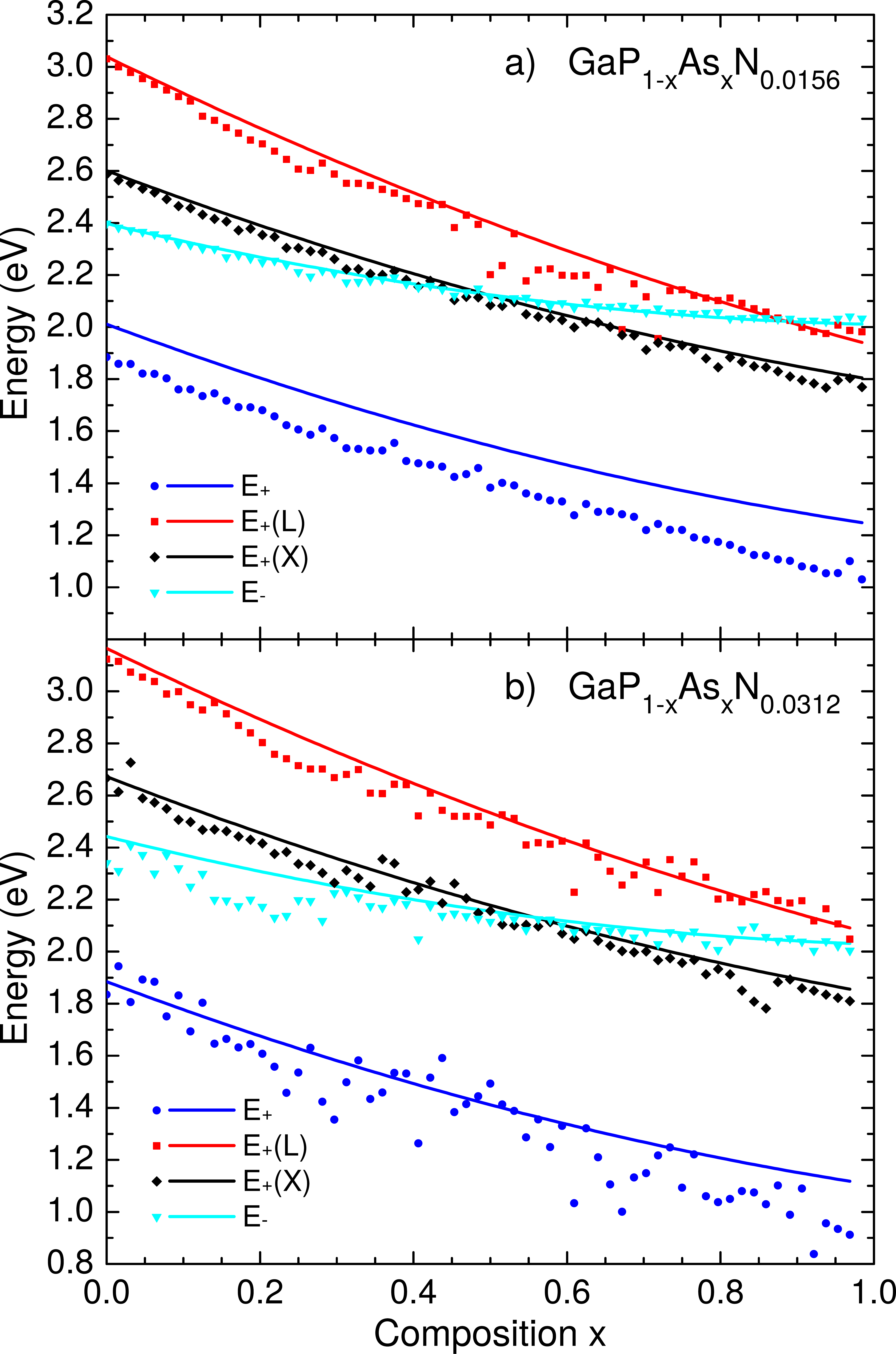}
\caption{Direct ($\Gamma$-$E_-$ and $\Gamma$-$E_+$) and indirect ($\Gamma$-$E_+(X)$ and $\Gamma$-$E_+(L)$) band gaps in the {\gapasne} alloy as a function of As concentration, with the fixed nitrogen content of (a) 1.56\% and (b) 3.12\%. The points represent results from DFT calculations while the curves are fitted with Eq.~(\ref{vegard}) using parameters listed in Table~\ref{Table:Parameters}.}
\label{Fig:GaPAsN-gaps}
\end{figure}

A composition dependence of direct and indirect optical transitions in {\gapasn} alloys is presented in Fig.~\ref{Fig:GaPAsN-gaps}. A few very interesting findings can be extracted from this figure. Treating the $E_{-}$ band as the IB and $E_{+}$ band as the CB it is interesting to analyse the minimum of CB since {\gapasn} alloys with the energy minimum at the $\Gamma$ point are potential candidates for two colours emitters. For such emitters the radiative recombination takes place at the $\Gamma$ point of BZ from the CB to the VB and from the IB to the VB. The channel of recombination between the IB and the VB is expected in the whole range of {\gapasn} content since the minimum of IB is at the $\Gamma$ point. Electrons easily thermalize to this point and recombine radiatively with holes. For the CB the carrier relaxation process is not such obvious and strongly depends on the content of {\gapasn} alloy. For {\gapasn} alloy with low N concentration [Fig.~\ref{Fig:GaPAsN-gaps} (a)] the order of the $\Gamma$, X and L energy position is similar to this which is observed for {\gapas} host, see Fig.~\ref{GaPAs_x}. In this case only for As-rich {\gapasn} alloys the minimum of CB is at the $\Gamma$ point, but with the increase in N concentration this order in the CB are no longer present, see Fig. ~\ref{Fig:GaPAsN-gaps} (b). For {\gapasn} alloys with the energy minimum in $E_{+}$ band out of the $\Gamma$ point (i.e., at the X or L point) carriers thermalize out of the center of BZ and cannot recombine radiatively since holes thermalize to the $\Gamma$ point. Such conditions are present for P-rich {\gapasn} alloys. Moreover, with the increase in N concentration these conditions enhance and, thus, such alloys are not promising for two colour emitters since energy minima in the $E_{+}$ band appear out of the $\Gamma$ point.

The other feature extracted from Fig.~\ref{Fig:GaPAsN-gaps} is that the energy separation between the IB and the CB at the $\Gamma$ point changes very weakly with the content of {\gapas} host while the energy difference between the VB and the IB changes almost twice in this content range. Such features of the electronic band structure can be utilized in band gap engineering in semiconductor devices.

As will be shown later, the nitrogen-related states are highly localized and, hence, the energy eigenvalues in calculations are highly dependent on the nitrogen atom distribution. The 1.56\% N-composition corresponds to only one atom in a 128 atom supercell which creates an ordered distribution of nitrogen atoms within the host {\gapas}. This results in relatively smooth composition-dependent $E_g(x)$ curves, with a small scattering of the points caused only by alloy fluctuations of the host {\gapas} atoms in our calculations, see Fig.~\ref{Fig:GaPAsN-gaps} (a).

In the case of 3.12\% N-content, there are two atoms in the supercell distributed by the SQS algorithm together with the host {\gapas} atoms. A variation of the distance between the two N-atoms results in a much higher scattering of the points, see Fig.~\ref{Fig:GaPAsN-gaps} (b). The scattering of data for $E_-$ in Fig.~\ref{Fig:GaPAsN-gaps} (b) provides an estimate of 0.1~eV for the energy scale of inhomogeneous broadening due to N-N configurational fluctuations. Limitations imposed by the size of the supercell and dilute N-content preclude us from having triplets and higher order N-clusters. Those details can be found elsewhere \cite{Kent_APL_79_2001}.

In order to provide an easy way of estimating the energy of $\Gamma-E_{\pm}$ transitions as well as $E_{+}$ energies in L and X points for different compositions of the {\gapasn} alloy with low nitrogen content ($x$ up to around $0.10$ and $T=0$~K) we have combined all of the results of the changes in energies as a function of composition and derived an approximate general formulas for $E_\pm$. The expression is based on  Eq.~(\ref{vegard}) for {\gapas}, where the values of $E_g^{AB}$ and $E_g^{AC}$ are calculated with the use of Eq.~(\ref{Eq:H-BAC}) for {\gapn} and {\gaasn} respectively. The $C_{NM}$ and $E_N$ parameters were linearly interpolated between {\gapn} and {\gapas} (see Table~\ref{Table:Parameters}). It is worth noticing, that although the $C_{NM}(\Gamma)$ and $E_N$ parameters for both {\gapn} and {\gaasn} can be extracted from the results, well established and commonly used experimental parameters have been used \cite{gapn_param,gaasn_param}. The parameters outside of the $\Gamma$ point, however, have not been determined before, therefore the values extracted from our calculations are used, creating the possibility for a complete description of the band edge energies within the whole BZ.

\begin{figure}[H]
\centering
\includegraphics[width=0.45\textwidth]{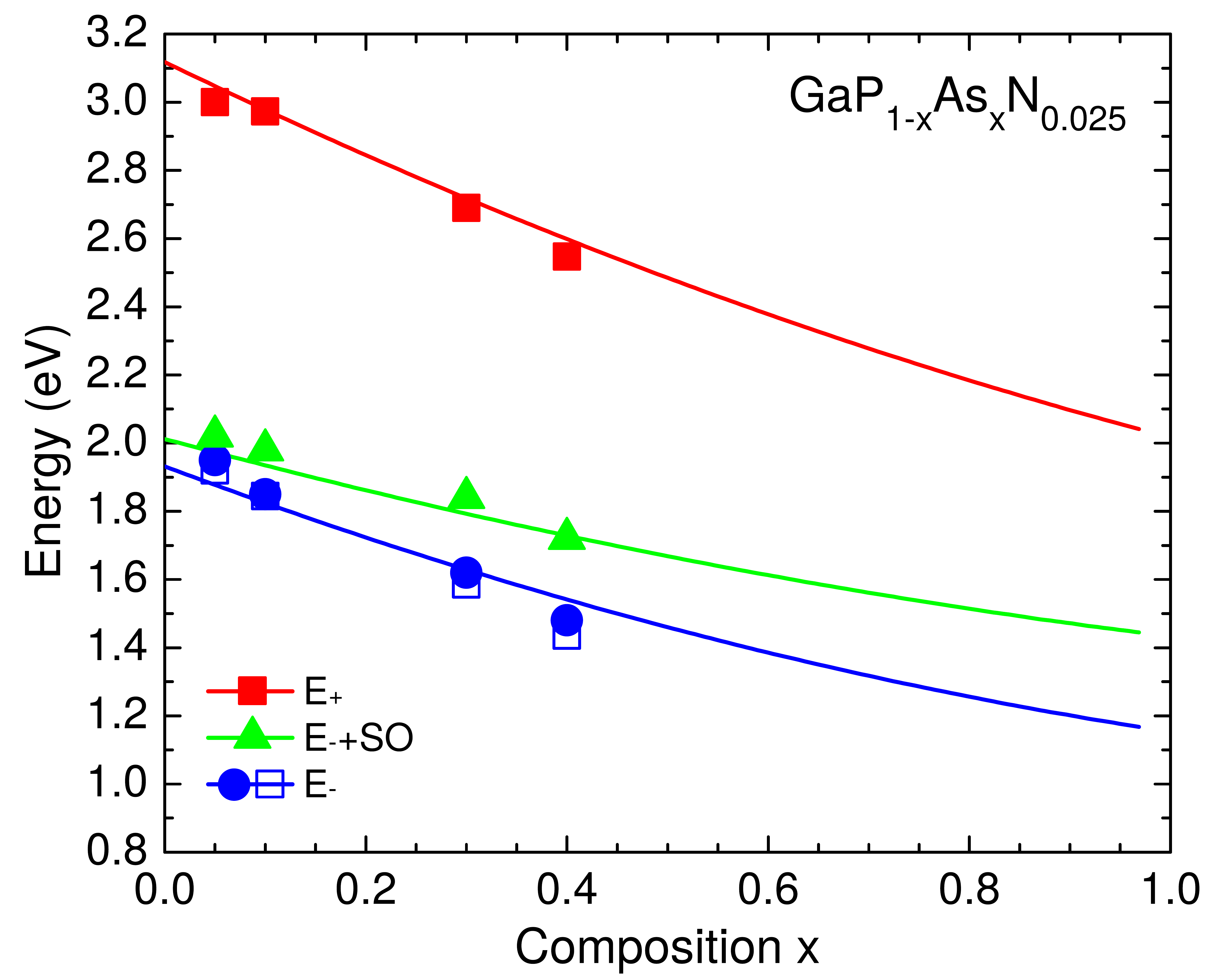}
\caption{Direct band gap transitions $\Gamma-E_-$, $\Gamma-E_+$, and the split-off band transition ($\Gamma_\text{SO}-E_-$) in the {\gapasn} alloy as a function of As concentration. Solid points represent modulated transmittance results, open points are results of photoreflectance spectroscopy, both from Ref.~\cite{Zelazna2017} where the nitrogen content has been determined to be around 2.5\%, although with high uncertainty. The lines are the fitted curves from our calculations (Eq.~(\ref{bac_vegard}) with parameters from Table~\ref{Table:Parameters}).}
\label{Fig:DFT_vs_EXP}
\end{figure}

We validate the model equations by comparing calculated energies of direct optical transitions in the {\gapasn} alloy with those determined experimentally. Figure ~\ref{Fig:DFT_vs_EXP} shows experimental points from Ref. \cite{Zelazna2017} obtained at room temperature as a function of As concentration together with our theoretical predictions for N = 2.5 \% and shifted by 100 meV, which corresponds to the temperature shift of band gap in this alloy between 0 and 300 K. As seen in Fig.~\ref{Fig:DFT_vs_EXP} the agreement between experimental data and our predictions is very good for both the $E_-$ and $E_+$ transition. It confirms that our calculations are reliable and satisfactorily describe the electronic band structure for this alloy. 

In a search for a gap between the IB and the host conduction band, we inspected the DOS of {\gapasn} alloys with low As and N compositions (Fig.~\ref{Fig:gapnas_dos}). It is difficult to infer directly from the DOS plots whether the gap would exists for more dilute compositions than those studied here (1.56\% of N), but the observed trends suggest that it would. Figure~\ref{Fig:gapnas_dos} suggests the gap most likely to appear in the alloy containing near 1.56\% of N and 3.12\% of As (solid red line), because the DOS between the two bands gets to a very low value, yet without a visible flat, zero density region. Increasing nitrogen concentration to 3.12\% clearly eliminates the gap. The energies of $\Gamma_v-E_-$ and $\Gamma_v-E_+$ transitions in GaP$_{0.9532}$As$_{0.0312}$N$_{0.0156}$ are approximately 1.8 and 2.7~eV, respectively, see Fig.~\ref{Fig:GaPAsN-gaps}(a). The ratio of band gaps (67\%) is almost ideal for the IBSC application \cite{Luque_NP_6_2012}. However, the full gap is 0.75~eV above the recommended value of 1.95~eV \cite{Luque_NP_6_2012}. As a result, increased losses due to transmittance of photons with energies below the band gap can negate benefits introduced by the IB. On the other hand this absorber can be integrated in tandem solar cells with the well developed Si-solar cells especially that the lattice mismatch between Si and {\gapasn} in this content regime is almost negligible. Therefore, we are fully convinced that {\gapasn} alloy with low N and As concentration is very promising for growth and exploration in the context of solar cell applications. 

\subsection{Carrier localization}\label{Subsec:Carrier localization}

\begin{figure*}
\centering
\includegraphics[width=0.95\textwidth]{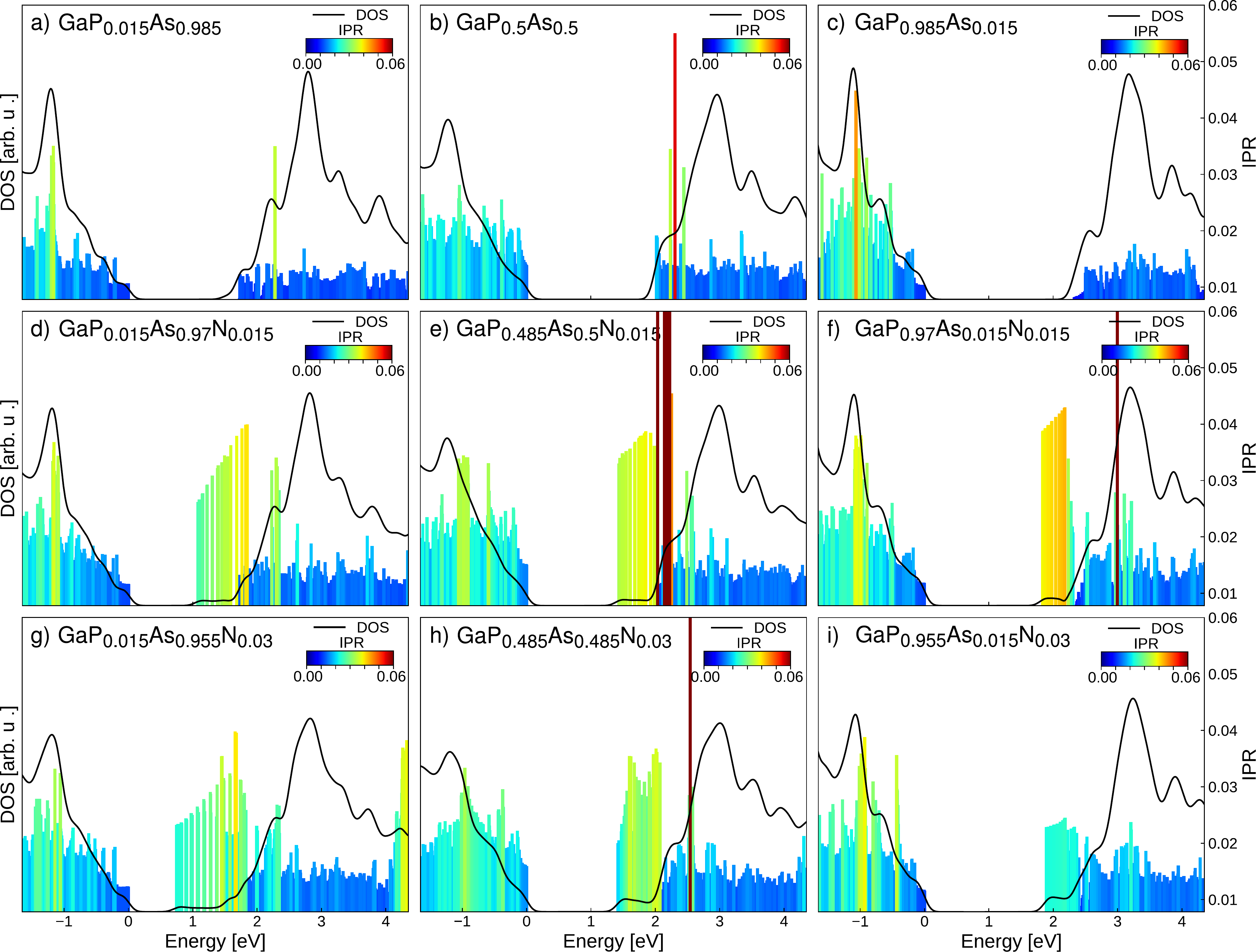}
\caption{Density of states (DOS) and spectrally-resolved inverse participation ratio (IPR) for states in vicinity of the band edges.}
\label{Fig:IPR}
\end{figure*}

Spatial localization of electronic states that participate in transport of photogenerated excitations is not desirable and should be assessed when considering new solar cell absorber materials.  The fact that incorporation of nitrogen into either GaP or GaAs causes a carrier localization is well known \cite{BUYANOVA2003467, RK_localization, RK_PRPL}. However, whether the localization changes with different composition of {\gapas} alloy as a host has not been discussed so far and only a few papers study the carrier localization in {\gapasn} \cite{baranowski_jap, woscholski_jap}.

Here we employ the inverse participation ratio (IPR) as a measure of localization. The IPR for each eigenvalue $E_i$ has been calculated based on the probabilities $p_n$ of finding an electron within a muffin-tin sphere of an atom $n$ \cite{Pashartis2017}
\begin{equation}\label{Eq:IPR}
\text{IPR}(E_i)=\frac{\sum_np_n^2(E_i)}{\left[ \sum_np_n(E_i)\right]^2}.
\end{equation}
Here the summation index $n$ runs over all atoms in the supercell. Hence, for a 128 atom supercell the IPR may span from $1/128$ for an eigenvalue with equal probabilities within the spheres of all atoms, to $1$ for an eigenvalue completely confined within a single atomic sphere. More details on the IPR as a measure of localization can be found in \cite{Pashartis2017}.

The IPR has been calculated for alloys with high P, high As, and nearly 50\% of As/P, all with with 1.56 and 3.12\% of N (Fig.~\ref{Fig:IPR}). The nitrogen-free host {\gapas} alloy has been also calculated for comparison. Further discussion concerning localization will focus on states in vicinity of the band edges  that are relevant for transport and optical properties.

States at the band edges of the nitrogen-free {\gapas} alloy can be classified as extended since they exhibit a very low IPR [Fig.~\ref{Fig:IPR} (a), (b), and (c)]. The compositional disorder mostly affects states located energetically deeper in the VB. The localization-free band edges are expected since this material system exhibits narrow low-temperature PL spectra \cite{Lai_PRL_44_1980,Oueslatii_JPCM_1_1989,Fried_PRB_39_1989}.

\begin{figure}[H]
\centering
\includegraphics[width=0.45\textwidth]{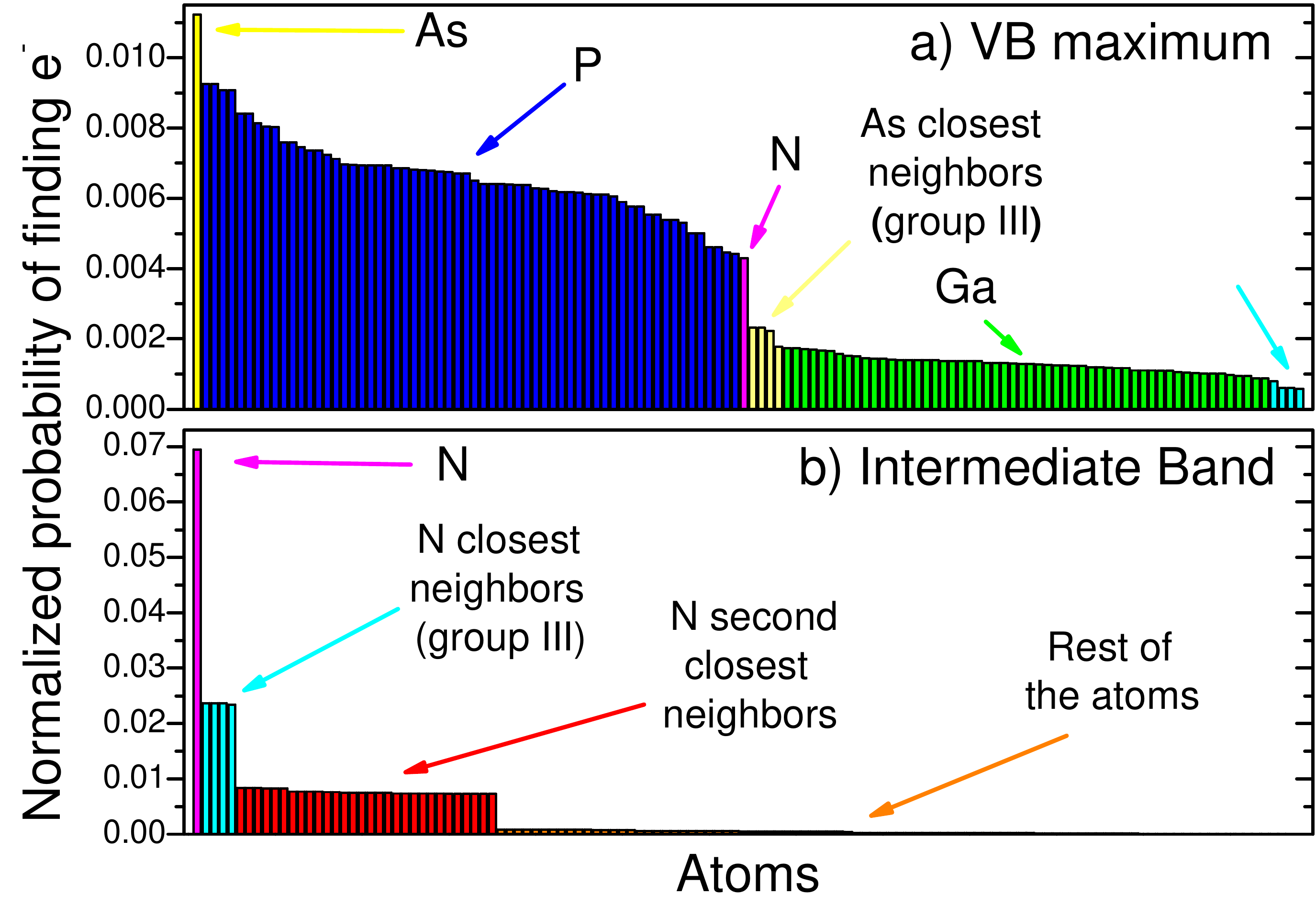}
\caption{The localization of carriers in GaP$_{0.9688}$As$_{0.0156}$N$_{0.0156}$ represented as a normalized probability of finding an electron within the sphere of each atom in descending order from left to right. Probabilities for (a) the top of the valence band and (b) the bottom of the intermediate band.}
\label{GaPAsN_local}
\end{figure}

The situation changes drastically with the introduction of nitrogen into the system, see Fig.~\ref{Fig:IPR} (d)-(i). In the As-rich limit [Fig.~\ref{Fig:IPR} (d)], nitrogen introduces localized states at the CB edge which corresponds to the energies of the IB. The corresponding orbitals are confined within the second nearest neighbor distance from the nitrogen atoms,  see Fig.~\ref{GaPAsN_local}.  The increase in nitrogen concentration [Fig.~\ref{Fig:IPR} (g)] does not influence the overall character of localization; the localized states only slightly spread into higher energies in the CB.

The P-rich {\gapasn} alloys behaves somewhat differently. The introduction of nitrogen still influences primarily the CB, with even stronger localization. The localized states spread over a smaller energy range near the band edge [Fig.~\ref{Fig:IPR} (f) corresponding to the IB]. However, the localization in the IB reduces significantly with the increase in nitrogen content [Fig. \ref{Fig:IPR} (i)] without any relevant influence on states at the VBE.

\section{Summary}
State-of-the-art density functional methods have been used to study the electronic band structure of nitrogen diluted {\gapasn} solid solutions to investigate formation of the intermediate band, direct and indirect optical transitions and localization of electronic states. The obtained results have been then confronted with experimental studies of these alloys, further confirming their reliability. The influence of introducing nitrogen to {\gapas} host on its electronic band structure has been discussed in terms of its use in optoelectronic devices such as intermediate band solar cells and light emitters. As expected, relatively small amounts of nitrogen introduced to {\gapas} allowed for an intermediate nitrogen-related band to appear below the conduction band, but the separation between the intermediate and conduction band may be present for low nitrogen concentrations in P-rich alloys only. Therefore such alloys have been chosen as potential candidates for absorbers suitable for tunable intermediate band solar cells, potentially as the second absorber in Si-based tandem solar cells. In addition, {\gapasn} alloys with the global minimum in conduction and intermediate band at the $\Gamma$ point of Brillouin zone has been identified in our calculations and proposed as potential candidate from group III-V for two colours emitters. For easy identification of the content of {\gapasn} with the desired electronic band structure interpolation formulas have been derived. It has been shown that the electronic band structure of {\gapasn} is well approximated with the electronic band structures of the highly mismatched ternary dilute nitrides [{\gapn} and {\gaasn} in this case] with the same nitrogen concentration used as hosts and then interpolated with the parabolic Vegard equation, commonly used for ordinary semiconductor alloys. In order to improve this interpolation by extending it to the entire Brillouin zone, necessary parameters have been determined and presented in this work. In addition the magnitude of carrier localisation has been calculated, presented as inverse participation ratio and discussed, revealing the highly localized nitrogen states in the intermediate band.

\section{Acknowledgments}
This work was supported within the grant of the National Science Centre Poland HARMONIA 2013/10/M/ST3/00638.
In addition, M. P. Polak acknowledges the financial support within the MNiSzW ``Diamond Grant" no. DI2013 006143.
O.R. would like to acknowledge funding provided by the Natural Sciences and Engineering Research Council of Canada under the Discovery Grant Programs RGPIN-2015-04518.
All DFT calculations have been performed in Wroclaw Centre of Networking and Supercomputing.

\bibliography{bib}

\end{document}